\documentclass[12pt]{article}

\usepackage{amsmath}
\usepackage{amssymb}
\usepackage{graphicx}

\newtheorem{theorem}{Theorem}

\newcommand{\cqfd}{\mbox{}\hfill\rule{.8em}{1.8ex}}
\newcommand{\tdx}{\tilde x}
\newcommand{\tdc}{\tilde c}
\newcommand{\tdcb}{\tilde{\bf c}}
\newcommand{\tdS}{\widetilde S}
\newcommand{\tddS}{\widetilde{\widetilde S}}

\title{On central tendency and dispersion measures for intervals and hypercubes}

\author{
Marie Chavent$^1$ and J\'er\^ome Saracco$^{1,2}$\footnote{preprint submitted to Communications in Statistics -
Theory and Methods}\\
\\
{\small $^1$ Universit\'e Bordeaux 1}\\
{\small Institut de Math\'ematiques de Bordeaux, UMR CNRS 5251}\\
{\small 351 cours de la lib\'eration, 33405 Talence Cedex, France}\\
{\small e-mail: \{Marie.chavent,Jerome.Saracco\}@math.u-bordeaux1.fr} \\
{\small $^2$ GREThA, UMR CNRS 5113} \\
{\small Universit\'e Montesquieu - Bordeaux IV}\\
{\small Avenue L\'eon Duguit, 33608 Pessac Cedex, France}}

\date{}

\begin{document}
\maketitle

\begin{abstract}
The uncertainty or the variability of the data may be treated by considering, rather than a single value for each data, the interval of values in which it may fall. This paper studies the derivation of basic description statistics for interval-valued datasets. We propose a  geometrical approach in the determination of summary statistics (central tendency and dispersion measures) for interval-valued variables.

\medskip

\end{abstract}

\section{Introduction}
In descriptive statistics, summary statistics are used to synthesize a set of real observations.
They usually involve:
\begin{itemize}
\item[-] a measure of location or central tendency, such as the arithmetic mean, median, interquartile mean or midrange,
\item[-] a measure of dispersion like the standard deviation, range, interquartile range or absolute deviation.
\end{itemize}

In this paper, we focus on obtaining basic descriptive statistics as central tendency and dispersion measures for interval-valued data.
Such data are often met in practice, they typically reflect the variability and/or uncertainty that underly the observed measurement. Interval data is a special case of `symbolic data', which also comprises
 set-valued categorical and quantitative variables as described, e.g., in Bock and Diday (2000).

Empirical extensions of summary statistics to the calculation of the mean and variance for interval valued-data have been given by Bertrand and Goupil (2000) and for histogram-valued data by Billard and Diday (2003).

In this paper, we propose a geometrical determination of summary statistics (mean, median, variance, absolute deviation,....) for interval-valued variables. This approach mimics the case of real-valued variables, with the absolute value of the difference between two real numbers being replaced by a distance between two intervals.

For real-valued variables, a geometrical way for defining a central value $c$ of a set $\{x_1,x_2,....,x_n\}$ of $n$ real observations is to choose $c \in \mathbb{R}$ as close as possible to all the $x_i$'s. Let us define the function $S_p$:

\begin{equation}
S_p(c) =\parallel {\bf x}- {\bf c} \parallel_p=\left\{
\begin{array}{ll}
(\sum_{i=1}^n \mid x_i-c \mid^p)^{1/p} & \mbox{for } p < \infty, \\
\max_{i=1\ldots n} \mid x_i-c \mid & \mbox{for } p=\infty,
\end{array}
\right. \label{eq:sdev}
\end{equation}

\noindent where   ${\bf x} \in \mathbb{R}^n$ is the vector of the $n$ observations $x_i$, $\parallel \cdot \parallel_p$ is the $L_p$ norm on $\mathbb{R}^n$, and ${\bf c}=c\mathbb{I}_n$
with  $\mathbb{I}_n$ the unit vector. Then one can use
\begin{equation}
\hat{c}=\arg \min_{c\in\mathbb{R}} S_p(c) ,
\end{equation}
as a central value and $S_p(\hat{c})$ as the associated dispersion measure. The above minimization problem has an explicit solution for $p=1,2,\infty$.
\begin{itemize}
\item When $p=1$,  the central value is $\hat{c}=x_M$ (the sample median)
and the corresponding dispersion is $S_1(x_M)=\sum_{i=1}^n \mid x_i-x_M \mid=n s_M$  where $s_M$ is the average absolute deviation from the median.
\item When $p=2$,  the central value is $\hat{c}=\bar{x}$ (the sample mean) and the corresponding dispersion is $S_2(\bar{x})=\sqrt{\sum_{i=1}^n (x_i-\bar{x})^2}=\sqrt{(n-1)}s$ where $s$ is the sample standard deviation.
\item When $p=\infty$,  the central value is $\hat{c}=x_R$ (the midrange) and the corresponding dispersion is $S_{\infty}(x_R)=\max_{i=1\ldots n} \mid x_i-x_R \mid=\frac{1}{2}w$ where $w$ is the sample range.
\end{itemize}
The pairs $(\bar{x}, s^2)$, $(x_M, s_M)$ and $(x_r, w)$  are then consistent with the use of respectively the $L_1$, $L_2$ and $L_\infty$ norms in the function $S_p$.

For interval-valued variables, we will use  the above geometrical approach to define coherent measures of central tendency and dispersion of a set  $\{\tdx_1,\tdx_2,....,\tdx_n\}$  of $n$ intervals $\tdx_i= [a_i,b_i] \in I=\{[a,b] \; |  \; a,b \in \mathbb{R} \; ,  \;   a \leq b \}$. A measure of central tendency $\tdc$ is now an interval $\tdc=[\alpha, \beta]$ defined in order to be as close as possible to all the $\tdx_i$'s. Replacing in (\ref{eq:sdev}) the terms $\mid x_i-c \mid $ by a distance  $d(\tdx_i,\tdc)$ between two intervals leads to the function $\tdS_{p}$ defined by:

\begin{equation}
\tdS_p(\tdc)=\left\{
\begin{array}{ll}
(\sum_{i=1}^n d(\tdx_i,\tdc)^p)^{1/p} & \mbox{for } p < \infty, \\
\max_{i=1\ldots n} d(\tdx_i,\tdc) & \mbox{for } p=\infty.
\end{array}
\right.
 \label{eq:s}
\end{equation}
The central interval $\hat{\tdc}=[\hat{\alpha}, \hat{\beta}]$ is then defined as
\begin{equation}
\hat{\tdc}=\arg \min_{\tdc\in I} \tdS_p(\tdc),
\end{equation}
and the corresponding dispersion measure is $\tdS_p(\hat{\tdc})$.

In the following, after a brief recall of some definitions of distances between intervals (section \ref{sec:dist}), we exhibit in section \ref{classic} particular cases of value $p$ and distance $d$ for which explicit formula of the lower and upper bounds of central intervals  $\hat{\tdc}$ have already been developed. Then we resolve in section \ref{main} the case where $p=2$ and $d$ is the Hausdorff distance and we show how  the corresponding central interval can be computed in a finite number of operations proportional to $n^3$.  We generalize in section \ref{gen} all these results to hypercubes. Finally, concluding remarks are given in section \ref{conclu}.

\section{Distances between intervals} \label{sec:dist}

Many distances between intervals have been proposed. They vary from simple ones to the more elaborated ones.  Elaborated distances taking into account both range and position have been  proposed in the framework of symbolic data analysis (see for instance, Chapter 8 and 11.2.2 of Bock and Diday, 2000, De Carvalho, 1998, Ichino and Yaguchi, 1994). Simple distances commonly used to compare $\tdx_1=[a_1,b_1]$ and $\tdx_2=[a_2, b_2]$ are the $L_p$ distances between:
\begin{itemize}
\item the two vectors $\left( \begin{array}{c} a_1 \\ b_1 \end{array} \right)$ and $\left( \begin{array}{c} a_2 \\ b_2 \end{array} \right)$ of the lower and upper bounds,
\item or the two vectors $\left( \begin{array}{c} m_1 \\ l_1 \end{array} \right)$ and $\left( \begin{array}{c} m_2 \\ l_2 \end{array} \right)$ of the midpoints $\displaystyle m_i=\frac{a_i+b_i}{2}$ and the half-lengths $\displaystyle l_i=\frac{b_i-a_i}{2}$.
\end{itemize}

General distances between sets like the Hausdorff distance (see Nadler, 1978), can also be used to compare two intervals. In the case of two intervals $\tdx_1=[a_1,b_1]$ and $\tdx_2=[a_2, b_2]$, the Hausdorff distance has the property to simplify to:
\begin{equation}
d(\tdx_1,\tdx_2)=\max(\mid a_1-a_2\mid ,\mid b_1-b_2 \mid ) \ . \label{eq:h1}
\end{equation}

By replacing in (\ref{eq:h1}) the lower bound $a_i$ by $(m_i-l_i)$ and the upper bound $b_i$ by $(m_i+l_i)$, and according to the following property defined for  $x$ and $y$ in $\mathbb{R}$,
$$
\max(|x-y|,|x+y|)=|x|+|y|,
$$
one can show that the Hausdorff distance can be written as:
\begin{equation}
d([a_1,b_1],[a_2, b_2])= \mid m_1-m_2\mid +\mid l_1-l_2\mid. \label{eq:h2}
\end{equation}

The Hausdorff distance between intervals has then the interesting property to be, at the same time,
\begin{itemize}
\item[-] a distance between sets,
\item[-] equal  to the $L_\infty$ distance  between the vectors $\left( \begin{array}{c} a_1 \\ b_1 \end{array} \right)$ and $\left( \begin{array}{c} a_2 \\ b_2 \end{array} \right)$,
\item[-] equal to the $L_1$ distance between the vectors  $\left( \begin{array}{c} m_1 \\ l_1 \end{array} \right)$ and $\left( \begin{array}{c} m_2 \\ l_2 \end{array} \right)$.
\end{itemize}

\section{Existing results on central intervals} \label{classic}
Explicit formula of the central interval $\hat{\tdc}=[\hat{\alpha}, \hat{\beta}]=\arg \min_{\tdc\in I} \tdS_p(\tdc)$ can be found in some particular cases. We remind these results already obtained and used in previous works (see for instance Chavent and Lechevallier, 2002, Chavent, 2004, De Carvalho et al., 2006).

\subsection{$L_1$ combination of Hausdorff distances}
When $p=1$ and $d$ is the Hausdorff distance, $\tdS_p(\tdc)$ reads:
\begin{equation}
\tdS_1(\tdc)=\sum_{i=1}^n( \mid m_i-\mu\mid +\mid l_i-\lambda\mid) \ , \label{eq:sss}
\end{equation}
where $\mu$ and $\lambda$ are the midpoint and the half-length of $\tdc=[\alpha,\beta]$.

\noindent Minimization of $\tdS_{1}(\tdc)$ boils down to the two minimization problems:
$$
\min_{\mu \in \mathbb{R}} \sum_{i=1}^n |m_i-\mu| \; \; \mbox{   and   } \; \;  \min_{ \lambda \in \mathbb{R}}  \sum_{i=1}^n |l_i-\lambda|.
$$

\begin{theorem}
In case of an $L_1$ combination of Hausdorff distances, the midpoint $\hat{\mu}$ and the half-length $\hat{\lambda}$ of the central interval $\hat{\tdc}$ are:
\begin{equation}
\hat{\mu}=median\{m_i \; |  \;  i=1,\ldots, n \}, \hspace{0.3cm}
\hat{\lambda}=median\{l_i \; |  \;  i=1,\ldots, n\}.
 \label{eq:med}
\end{equation}

\end{theorem}

\subsection{ $L_\infty$ combination of Hausdorff distances}
When $p=\infty$ and $d$ is the Hausdorff distance, $\tdS_p(\tdc)$ reads:
\begin{equation}
\tdS_{\infty}(\tdc)=\max_{i=1,\dots,n}\max\Big\{\mid a_i-\alpha \mid ,\mid b_i-\beta \mid\Big\} \ , \label{eq:ra}
\end{equation}
i.e.
$$
\tdS_{\infty}(\tdc)=\max\Big\{ \max_{i=1,\dots, n} \mid a_i-\alpha \mid \, , \max_{i=1,\dots,n}\mid b_i-\beta \mid\Big\} \ .
$$

\noindent Minimization of  $\tdS_{\infty}(\tdc)$ boils down to the two minimization problems:
$$
\min_{\alpha \in \mathbb{R}} \max_{i=1,\ldots, n}|a_i-\alpha|  \ \ \mbox{ and } \
\min_{\beta \in \mathbb{R}} \max_{i=1,\ldots, n}|b_i-\beta| \ .
$$
\begin{theorem}
In case of an $L_\infty$ combination of Hausdorff distances, the lower bound $\hat{\alpha}$ and the upper bound $\hat{\beta}$ of the central interval $\hat{\tilde{c}}$  are:
\begin{equation}
\hat{\alpha}=\displaystyle \frac{a_{(n)}-a_{(1)}}{2}, \hspace{0.3cm}
\hat{\beta}=\displaystyle \frac{b_{(n)}-b_{(1)}}{2},
 \label{eq:sol4}
\end{equation}
where $a_{(n)}$ (resp. $b_{(n)}$) is the largest lower bound (resp. upper bound) and $a_{(1)}$ (resp. $b_{(1)}$) is the smallest lower bound (resp. upper bound).
\end{theorem}

\subsection{ $L_2$ combination of $L_2$ distances}
For $p=2$, an explicit solution is easily defined when $d$ is the $L_2$ distance between either the middles and half lengths of the intervals or between their lower and upper bounds. For instance in the first case, $\tdS_p(\tdc)$ reads:
\begin{equation}
\tdS_2(\tdc)=\sqrt{\sum_{i=1}^n d(\tdx_i,\tdc))^2}=\sqrt{\sum_{i=1}^n (\mid m_i-\mu\mid)^2 +(\mid l_i-\lambda\mid)^2} \ . \label{eq:s2}
\end{equation}

\begin{theorem}
In case of an $L_2$ combination of $L_2$ distances between midpoints and half lengths, the midpoint $\hat{\mu}$ and the half-length $\hat{\lambda}$ of the central interval $\hat{\tdc}$  are:
$$
\hat{\mu}=\frac{1}{n} \sum_{i=1}^n m_i \  \  \mbox{and  } \ \  \hat{\lambda}=\frac{1}{n} \sum_{i=1}^n l_i \ .
$$
In case of an $L_2$ combination of $L_2$ distances between lower and the upper bounds, the lower and upper bounds of the intervals of the central interval $\hat{\tdc}$ are:
$$
\hat{\alpha}=\frac{1}{n} \sum_{i=1}^n a_i \  \  \mbox{and  } \ \  \hat{\beta}=\frac{1}{n} \sum_{i=1}^n b_i \ .
$$
\end{theorem}

\section{Main result} \label{main}
We study here the case of an $L_2$ combination of Hausdorff distances. When $p=2$ and $d$ is the Hausdorff distance, $\tdS_p(\tdc)$ reads: \begin{equation}
\label{eq: l2 norm via eq:h1}
\left(\tdS_2(\tdc)\right)^2=\sum_{i=1}^n (\max(\mid a_i-\alpha \mid ,\mid b_i-\beta \mid)^2  \ .
\end{equation}

\begin{theorem}
In case of an $L_2$ combination of Hausdorff distances, the central interval $\tdc$ which minimizes (\ref{eq: l2 norm via eq:h1}) can be computed in a finite number of operations proportional to $n^3$.
\end{theorem}
{\it Proof: }
The square is an increasing function over positive numbers, so formula (\ref{eq: l2 norm via eq:h1}) can be rewritten:
\begin{equation}
\label{eq: l2 norm via eq:h1-2}
\left(\tdS_2(\tdc)\right)^2=\sum_{i=1}^n \max\Big( (a_i-\alpha)^2 , (b_i-\beta)^2 \Big)  \ .
\end{equation}
On the other hand, using midpoints and half-lengths, one obtains:
$$
(a_i-\alpha)^2 - (b_i-\beta)^2 = -4(m_{i}-\mu)(l_{i}-\lambda) \ .
$$
So we see that the maximum in (\ref{eq: l2 norm via eq:h1-2}) is $( a_i-\alpha)^2$ if $(m_{i}-\mu)(l_{i}-\lambda)\leq0$, and $( b_i-\beta)^2$ if $(m_{i}-\mu)(l_{i}-\lambda)\geq0$.

Let us denote by $(m_{(1)}, \dots, m_{(n)})$, resp. $(l_{(1)}, \dots, l_{(n)})$, the sample of the midpoints, resp. the half-lengths,  organized in increasing order. Let us define the intervals:
\begin{equation}
\label{L&M}
\hspace{-0,6em}
\begin{array}{l}
 M_{j}=[m_{(j)},m_{(j+1)}] ,\ \ j=0, \dots, n,\\
 L_{k}=[l_{(k)},l_{(k+1)}] ,\ \ k=0, \dots, n ,
 \end{array}
 \hspace{-1,3em}
\end{equation}
with $m_{(0)}=l_{(0)}=-\infty$ and $m_{(n+1)}=l_{(n+1)}=+\infty$.
For all $(\mu, \lambda)$ in any rectangle $Q_{j,k}=M_{j} \times L_{k}$, the product $(m_{i}-\mu)(l_{i}-\lambda)$ has a given sign, for each $i=1, \dots, n$. So the formula (\ref{eq: l2 norm via eq:h1-2}) for $\left(\tdS_2(\tdc)\right)^2$ simplifies over such a rectangle to:
\begin{equation}
\label{eq:Sjk}
  \tdS_{j,k}(\tdc)=\sum_{i \in I_{a,j,k}} (a_i-\alpha)^2 + \sum_{i \in I_{b,j,k}} (b_i-\beta)^2 \ ,
\end{equation}
where:
\begin{equation}
\label{Iajk}
  I_{a,j,k}=\big\{ i \in \{1\dots n\} | \big(m_{i} -\frac{m_{(j)}+m_{(j+1)}}{2}\big)\big(l_{i}-\frac{l_{(k)}+l_{(k+1)}}{2} \big) \leq 0 \big\},
\end{equation}
\begin{equation}
\label{Ibjk}
  I_{b,j,k}=\big\{ i \in \{1\dots n\} | \big(m_{i} -\frac{m_{(j)}+m_{(j+1)}}{2}\big)\big(l_{i}-\frac{l_{(k)}+l_{(k+1)}}{2} \big) > 0 \big\}.
\end{equation}

Hence the minimization of $\left(\tdS_2(\tdc)\right)^2$ over $\mathbb{R}^2$ is equivalent to the resolution, for $j,k=0,1\dots n$, of the $(n+1)^2$ constrained quadratic problems:
\begin{equation}
\label{Pjk}
\mathrm{(P}_{j,k}\mathrm{)} \left\{
\begin{array}{ll}
    \mbox{find } (\alpha,\beta)=(\hat{\alpha}_{j,k},\hat{\beta}_{j,k}) \mbox{ which minimizes }  \tdS_{j,k}(\alpha,\beta) &    \\
    \mbox{under the constraints:} &   \\
    2m_{(j)} \leq \alpha+\beta \leq 2m_{(j+1)} \mbox{ and } 2 l_{(k)} \leq \beta-\alpha \leq 2l_{(k+1)}
\end{array} \right.
\end{equation}
whose resolution is described in the Appendix.

The central interval $\hat{\tdc}=[\hat{\alpha},\hat{\beta}]$ is then given by:
\begin{equation}
\label{eq: hat alpha beta}
  (\hat{\alpha},\hat{\beta})= \arg\min_{j,k=0,1 \dots, n} \tdS_{j,k}(\hat{\alpha}_{j,k},\hat{\beta}_{j,k}).
\end{equation}

Because the number of operations in the resolution of  (\ref{Pjk}) is proportional to $n$, the number of operations for the calculation of $(\hat{\alpha},\hat{\beta})$ is proportional to $n^3$.

\cqfd

\section{The multidimensional case} \label{gen}

We consider now a set of $n$ $k$-dimensional intervals $ \{{\bf \tdx}_1,\ldots , {\bf \tdx}_n\}$ with ${\bf \tdx}_i=[{\bf a}_i,{\bf b}_i]$ and ${\bf a}_i,{\bf b}_i \in \mathbb{R}^k$. A $k$-dimensional interval ${\bf \tdx}_i$ can also be viewed as a regular hyperparallelepiped ${\bf \tdx}_i=\prod_{j=1}^k \tdx_i^j$  with $\tdx_i^j=[a_i^j,b_i^j]$ where $a_i^j$ (resp. $b_i^j$) is the $j$th coordinate of ${\bf a}_i$ (resp. ${\bf b}_i$).  By misuse of language  the ${\bf \tdx}_i$'s will be called hypercubes in the rest of the paper.

The above geometrical approach can then be used to define a central hypercube (also called centrocube or prototype) of a set of $n$ hypercubes $ \{{\bf \tdx}_1,\ldots , {\bf \tdx}_n\}$, which is now a $k$-dimensional interval  ${\bf \tdc}=[\boldsymbol{\alpha},{\boldsymbol \beta}]$ with $\boldsymbol  \alpha$ and $\boldsymbol  \beta$ in $\mathbb{R}^k$. Replacing in (\ref{eq:s}) the terms $d(\tdx_i,\tdc)$ by a distance  $D({\bf \tdx}_i,{\bf \tdc})$ between two hypercubes leads to the function $\tddS_{p}$ defined by:

\begin{equation}
\tddS_p({\bf \tdc})=\left\{
\begin{array}{ll}
(\sum_{i=1}^n D({\bf \tdx}_i,{\bf \tdc})^p)^{1/p} & \mbox{for } p < \infty, \\
\max_{i=1\ldots n} D({\bf \tdx}_i,{\bf \tdc}) & \mbox{for } p=\infty.
\end{array}
\right.
 \label{eq:sk}
\end{equation}

The centrocube  ${\bf \tdc}=[\boldsymbol{\alpha},{\boldsymbol \beta}]$ is then be defined by
\begin{equation}
\hat{ \tdcb}=\arg \min_{\tdc\in I} \tddS_p({\bf \tdc}). \label{eq:cent}
\end{equation}

There exists many possible distances between hypercubes (see for instance Bock, 2002). Once again, depending on the distance $D$ and on the value $p$ in $\tddS_p( {\bf \tdc} )$, the centrocube is more or less difficult to calculate.

A first distance  $D$ that could be used is the Hausdorff distance between two hypercubes:
\begin{equation}
 D({\bf \tdx}_1,{\bf \tdx}_2)= \max(h({\bf \tdx}_1,{\bf \tdx}_2),h({\bf \tdx}_2,{\bf \tdx}_1))
\end{equation}
\noindent with
\begin{equation}
h({\bf \tdx}_1,{\bf \tdx}_2)= \sup_{a \in {\bf \tdx}_1} \inf_{b \in {\bf \tdx}_2} \delta(a,b) \label{eq:h}
\end{equation}
where $\delta$ is an arbitrary metric on $\mathbb{R}^k$.
We have seen that in the one-dimensional case, the Hausdorff distance simplifies to (\ref{eq:h1}) but the calculation of this distance for higher dimensions is more involved and depends of the choice of the metric $\delta$.  If $\delta$ is the Euclidean metric  for instance, there exist algorithms that compute the Hausdorff distance between two hypercubes in a finite number of steps (see e.g., Bock, 2005) but as far as we know, there exist no algorithm to compute the centrocube. If $\delta$ is the $L_\infty$ metric, an explicit solution of the centrocube exists when $p=\infty$ (see Chavent, 2004). In other cases, the definition of centrocubes for the original Hausdorff distance between hypercubes still remains a subject to investigate.

Another approach which makes explicit definitions of centrocubes easier to find, is to use a distance $D$  that is a combination of coordinate-wise one-dimensional interval distances $d$:

\begin{equation}
D({\bf \tdx}_1,{\bf \tdx}_2)=\left\{
\begin{array}{ll}
(\sum_{j=1}^k d(\tdx_1^j,\tdx_2^j)^q)^{1/q} & \mbox{for } q < \infty, \\
\max_{j=1\ldots k} d(\tdx_1^j,\tdx_2^j) & \mbox{for } q=\infty.
\end{array}
\right.
\label{eq:dist}
\end{equation}

When $p=q$, $\left(\tddS_p( {\bf \tdc})\right)^p$ reads:

\begin{equation}
 \left(\tddS_p({\bf \tdc})\right)^p= \sum_{i=1}^n \sum_{j=1}^k  \left(d(\tdx_i^j,\tdc^j)\right)^p \label{eq:cubes}
\end{equation}

Because $d(\tdx_i^j,\tdc^j) \geq 0$, it sufficient to find for each component $j$ the central interval $\hat{\tdc}^j$ which minimizes $\sum_{i=1}^n d(\tdx_i^j,\tdc^j)\ $, so that the centrocube is the product of the central intervals of each variable. The results presented in sections 3 and 4 concerning central intervals can then be applied directly to define this `coordinate-wise' centrocube.

\section{Concluding remarks} \label{conclu}
In this paper, we proposed different solutions for the determination of  central intervals and hypercubes.
These results have applications in clustering. Indeed, the existence of explicit formula for the computation of the centrocube is useful in dynamic clustering (see Diday and Simon, 1976), because it ensures the decreasing at each iteration of the criterion $\tddS_p$.  `Coordinate-wise' centrocubes have been defined as prototype in several dynamical clustering algorithms of interval data. The `coordinate-wise' centrocube for $p=q=1$ is used with the Hausdorff distance in Chavent and Lechevallier (2002)  and with the $L_1$ distance between the lower and the upper bounds in De Souza and De Carvalho (2004). The case $p=q=2$ is used by de Carvalho et al. (2006) with the $L_2$ distance between the lower and upper bounds. The algorithm proposed in section 4 for the determination of the central interval in the case of $L_2$ combination of Hausdorff distances gives a solution for the case $p=q=2$ and the Hausdorff distance.

Another application of these results concern the data scaling. Dealing with scalar variables measured on very different scales is already  a problem when comparing two objects globally on all the variables. For instance, the Euclidean distance or more generaly the $L_q$ distance will give more importance to variables of strong dispersion and the comparison between objects will only reflect their differences on those variables. A natural way to avoid this effect is to use a normalized distance. A $L_q$ normalized component-wise distance between hypercubes could then be:

\begin{equation}
D({\bf \tdx}_1,{\bf \tdx}_2)=\left\{
\begin{array}{ll}
(\sum_{j=1}^k (\frac{d(\tdx_1^j,\tdx_2^j)}{\tdS(\hat{\tdc}^j)})^q)^{1/q} & \mbox{for } q < \infty, \\
\max_{j=1\ldots k} \frac{d(\tdx_1^j,\tdx_2^j)}{\tdS(\hat{\tdc}^j)} & \mbox{for } q=\infty.
\end{array}
\right.
\label{eq:dintnor}
\end{equation}

\noindent where $\tdS(\hat{\tdc}^j)$ is the dispersion measure associated to a central interval $\hat{\tdc}^j$. For coherency reasons, it seems reasonable to use the same exponent ($q=p$):
\begin{itemize}
\item[-] to aggregate the intervals in the search of the central interval and the evaluation of the dispersion for each variable (exponent $p$ in (\ref{eq:s})),
 \item[-] and to evaluate the distance between objects (exponent $q$ in (\ref{eq:dintnor})).
\end{itemize}

To conclude, a natural extension of these results concerns weighted central tendency and dispersion measures. This point is currently under investigation.

\bigskip

\noindent
{\bf Acknowledgments}\\
The authors thank  G. Chavent  for his helpful contribution to the resolution of problem $\mathrm{(P}_{j,k}\mathrm{)}$ in the Appendix. They would like also to thank the associate editor and the reviewers for their useful comments.

\appendix
\section*{Appendix: Resolution of problem $\mathrm{(P}_{j,k}\mathrm{)}$}

We describe here the resolution of one of the minimization problems $\mathrm{(P}_{j,k}\mathrm{)}$ of equation (\ref{Pjk}). We drop the subscripts $j,k$, and we write $m_{-}$ instead of $m_{(j)}$, $m_{+}$ instead of $m_{(j+1)}$, $l_{-}$ instead of $l_{(j)}$ and $l_{+}$ instead of $l_{(j+1)}$. We use the midpoint and half-length variables $\mu = (\alpha + \beta)/2$ and $\lambda = (\beta-\alpha)/2$, and we denote by $Q$ the rectangle
\begin{equation}
\label{rect Q}
  Q=\{ (\mu,\lambda) \mbox{ such that } m_{-} \leq \mu \leq m_{+} \quad \mbox{and} \quad l_{-} \leq \lambda \leq l_{+} \} \ .
\end{equation}
With these notations, the problem to solve is now:
\begin{equation}
\label{Pappendix}
\mathrm{(P)}  \hspace{3em}  \mbox{find } (\hat{\mu},\hat{\lambda}) \mbox{ which minimizes }  \widetilde{S}(\mu,\lambda)
\mbox{ over } Q,  \hspace{2em}
\end{equation}
where the objective funtion is:
\begin{equation}
\label{eq:Sappendix}
  \widetilde{S}(\mu,\lambda)=\sum_{i \in I_{a}} (a_i-\mu + \lambda)^2 + \sum_{i \in I_{b}} (b_i-\mu - \lambda)^2,
\end{equation}
with $I_{a}$ and $I_{b}$ defined respectively in (\ref{Iajk}) and (\ref{Ibjk})
This objective function  is convex and quadratic (the level lines of $\widetilde{S}$ are - possibly degenerated - ellipses with axis parallel to the directions $\lambda = \mu$ and $\lambda = - \mu$), and the constraints  in (\ref{rect Q}) are linear, so that the resolution of $\mathrm{(P)}$ is equivalent to that of the associated Kuhn-Tucker system of necessary conditions.

We describe now the corresponding algorithm. We have eliminated the consideration of some dead-end cases by taking advantage of the convexity of the problem: when the solution $(\hat{\mu},\hat{\lambda})$ of  $\mathrm{(P)}$ is on one edge of $Q$ (possibly at a corner of $Q$) , the unconstrained minimizer $(\check{\mu},\check{\lambda})$ of $\widetilde{S}$ and the center of $Q$ are necessarily  on different sides of the line containing this edge.
Hence the edges of $Q$ which can possibly contain the solution $(\hat{\mu},\hat{\lambda})$ are those which contain the $L^2$-projection of $(\check{\mu},\check{\lambda})$ on $Q$.

We suppose for simplicity that the midpoints and half-length of all intervals are distinct:
\begin{equation}
\label{eq:strictly increasing} \left\{
\begin{array}{ccccccc}
      m_{(1)}&<&m_{(2)}&< &\dots& <&m_{(n)}    \\
      l_{(1)}&<&l_{(2)}&< &\dots &<&l_{(n)}
\end{array} \right.
\end{equation}

One computes first, in a loop from $i$ to $n$ over the samples:
\begin{equation}
\label{prel cal}
\left\{ \begin{array}{cclcccl}
 n_{a}& =& \sum_{i \in I_{a}} 1 & , &  n_{b} &=& \sum_{i \in I_{b}} 1 \,\ , \\
 A &=&\sum_{i \in I_{a}} a_{i}  & , &  B &= &\sum_{i \in I_{b}} b_{i} \ ,\\
 A_2& =& \sum_{i \in I_{a}} a_{i}^2 & , &  B_2 &=& \sum_{i \in I_{b}} b_{i}^2 \ ,
\end{array} \right.
\end{equation}
with the convention that the sum is zero if the set $I_{a}$ or $I_{b}$ of indices is empty. Notice that $n_{a}$ is the number of indices in $I_{a}$, and $n_{b}$ is the number of indices in $I_{b}$, so that  $n=n_{a}+n_{b}$.
With these notations, the gradient of $S$:
$$
\nabla \widetilde{S}(\mu,\lambda)=2 \left( \begin{array}{c}
- \sum_{i \in I_{a}} (a_i-\mu + \lambda) - \sum_{i \in I_{b}} (b_i-\mu - \lambda) \\
 + \sum_{i \in I_{a}} (a_i-\mu + \lambda) - \sum_{i \in I_{b}} (b_i-\mu - \lambda)  \end{array} \right).
$$
simplifies to:
\begin{equation}
\label{nabla}
\nabla \widetilde{S}(\mu,\lambda)=2 \left( \begin{array}{c}
- A - B+(n_{a}+n_{b})\mu - (n_{a}-n_{b})\lambda \\
+A - B-(n_{a}-n_{b})\mu +(n_{a}+n_{b})\lambda  \end{array} \right)\ .
\end{equation}

\vspace{1ex}

\noindent The minimizer $(\hat{\mu},\hat{\lambda})$ of problem $\mathrm{(P)}$ can be computed as follows:
\begin{enumerate}
  \item If $n_{a}=0$ (a similar reasoning can be done if $n_{b}=0$), then function $\widetilde{S}$ reduces over $Q$ to:
  $$
  \widetilde{S}(\mu,\lambda)= \sum_{i=1, \dots, n} (b_{i} - \mu - \lambda)^2 \ ,
  $$
and the level lines of $\widetilde{S}$ degenerate to the straight lines $\mu+\lambda=\mbox{constant}$. The unconstrained minimizers $(\check{\mu},\check{\lambda})$ of  $\widetilde{S}$ are then on the line:
$$
\mathrm{(L)} \quad \quad n(\mu+\lambda) = B \ .
$$
If the line  $\mathrm{(L)}$ goes through $Q$, problem $\mathrm{(P)}$ has an infinite number of solutions, with at least one of them (in general two) being on the boundary of $Q$. If $\mathrm{(L)}$ does not hit $Q$, the unique solution of  $\mathrm{(P)}$ is located at the corner of $Q$ closest to $\mathrm{(L)}$. In both cases, $\mathrm{(P)}$ admits at least one solution $(\hat{\mu},\hat{\lambda})$ on one edge of $Q$. If we denote by $Q^*$ the rectangle on the other side of this edge (for which $\tilde{n}_{a}=1\neq 0$), one sees that $(\hat{\mu},\hat{\lambda}) \in Q^*$, so that the minimum $\tilde{S}^*_{\mbox{\tiny min}}$ of $\tilde{S}$ over $Q^*$ will necessarily be smaller than $\tilde{S}_{\mbox{\tiny min}}$, the minimum of $\widetilde{S}$ over $Q$ (as $(\hat{\mu},\hat{\lambda}) \in Q^*)$. So there is no point in computing $\tilde{S}_{\mbox{\tiny min}}$, and we can skip the resolution of problem $\mathrm{(P)}$.

  \item  If $n_{a} > 0$ and $n_{b} > 0$, the unconstrained minimizer $(\check{\mu},\check{\lambda})$ of $\widetilde{S}$ is unique. It is given by:
  \begin{equation}
\label{uc optimizer} \left\{
\begin{array}{lcl}
     \sum_{i \in I_{a}} a_i =& n_{a}(\check{\mu}-\check{\lambda})  & =n_{a}\check{\alpha}, \\
     \sum_{i \in I_{b}} b_i =&  n_{b}(\check{\mu}+\check{\lambda})  & =n_{b}\check{\beta}.
\end{array} \right.
\end{equation}

If $(\check{\mu},\check{\lambda})\in Q$ then set $\hat{\mu}=\check{\mu}\ ,\ \hat{\lambda}=\check{\lambda}$, and problem is solved.

If not, go to the next step.

\item Compute the $L^2$-projection $(\check{\check{\mu}},\check{\check{\lambda}})$ of $(\check{\mu},\check{\lambda})$ on $Q$ :
  \begin{equation}
\label{proj}
\hspace{-1em}
\check{\check{\mu}} = \left\{
\begin{array}{lcrcl}
   m_{-}  & \mbox{if}&  \check{\mu} &\leq& m_{-}  \\
   \check{\mu} & \mbox{if}& m_{-}\hspace{-0,6em}&\leq &\check{\mu}  \leq m_{+} \\
   m_{+}  & \mbox{if}&  m_{+}\hspace{-0,6em} &\leq& \check{\mu}
\end{array}
\right. \ ,\
\check{\check{\lambda}} = \left\{
\begin{array}{lcrcl}
l_{-}  & \mbox{if}&  \check{\lambda} &\leq& l_{-}  \\
   \check{\lambda} & \mbox{if}& l_{-}\hspace{-0,6em}&\leq &\check{\lambda}  \leq l_{+} \\
   l_{+}  & \mbox{if}&  l_{+}\hspace{-0,6em} &\leq& \check{\lambda}
  \end{array}
\right.
\end{equation}
  \item If the projection is on a edge of $Q$, say for example $\check{\check{\mu}}=m_{-}\,,\, l_{-}< \check{\check{\lambda}} < l_{+}$ (left edge), determine $\check{\check{\check{\lambda}}}$ which zeroes the component of $\nabla \widetilde{S}$ along this edge (here the second component as the edge is parallel to the second axis $\mu = 0$):
  \begin{equation}
\label{eq:tttdl cote}
 + \sum_{i \in I_a} a_i-n_{a}(m_{-}\, - \check{\check{\check{\lambda}}}) - \sum_{i \in I_b} b_i+n_{b}(m_{-} + \check{\check{\check{\lambda}}}) =0.
\end{equation}
Then set:
\begin{equation}
\label{eq:solution cas cote}
\hat{\mu}=m_{-}\quad , \quad \hat{\lambda} = \left\{
\begin{array}{lcrcl}
   l_{-}  & \mbox{if}&  \check{\check{\check{\lambda}}} &\leq& l_{-} \ , \\
   \check{\check{\check{\lambda}}} & \mbox{if}& l_{-}\hspace{-0,6em}&\leq &\check{\check{\check{\lambda}}}  \leq l_{+} \ , \\
   l_{+}  & \mbox{if}&  m_{+}\hspace{-0,6em} &\leq& \check{\check{\check{\lambda}}} \ ,
\end{array}
\right. \ ,\
\end{equation}
and problem is solved.
 \item If the projection is at a corner of $Q$, say for example ${\check{\check{\mu}}}=m_{-}\,,\, \check{\check{\lambda}} = l_{-}$ (lower-left corner), evaluate the gradient $\nabla \widetilde{S}= (g_{\mu},g_{\lambda})$ at the corner.
 \begin{itemize}

 \item If $g_{\mu} \geq 0$ and $g_{\lambda} \geq 0$, set $\hat{\mu}=m_{-}\ ,\ \hat{\lambda}=l_{-}$, and problem is solved.

 \item If $g_{\mu} < 0$ and $g_{\lambda} \geq 0$, (the objective function is decreasing when one leaves the lower-left corner to the right on the lower edge of $Q$), determine $\check{\check{\check{\lambda}}}$ which zeroes the component of $\nabla \widetilde{S}$ along this edge (here the first component as the edge is parallel to the first axis $\lambda = 0$):
 \begin{equation}
\label{eq:tttdm coin 1}
- \sum_{i \in I_a} a_i+n_{a}(\check{\check{\check{\mu}}} - l_{-}) - \sum_{i \in I_b} b_i+n_{b}(\check{\check{\check{\mu}}} + l_{-}) =0.
\end{equation}
 Then set:
\begin{equation}
\label{eq:solution cas coin 1}
\hat{\mu}= \left\{
\begin{array}{lcrcl}
   \check{\check{\check{\mu}}} & \mbox{if}& m_{-}\hspace{-0,6em}&< &\check{\check{\check{\mu}}}  \leq m_{+} \ , \\
   m_{+}  & \mbox{if}&  m_{+}\hspace{-0,6em} &\leq& \check{\check{\check{\mu}}} \ ,
\end{array}
\quad , \quad \hat{\lambda}=l_{-} \ ,
\right.
\end{equation}
and problem is solved.
\item  If $g_{\mu} \geq 0$ and $g_{\lambda} < 0$, similarly determine $\check{\check{\check{\mu}}}$ which zeroes the component of $\nabla \widetilde{S}$ along the left edge of $Q$:
  \begin{equation}
\label{eq:tttdl coin 2}
+ \sum_{i \in I_a} a_i-n_{a}(m_{-}\, - \check{\check{\check{\lambda}}}) - \sum_{i \in I_b} b_i+n_{b}(m_{-}\, + \check{\check{\check{\lambda}}}) =0
\end{equation}
 Then set:
\begin{equation}
\label{eq:solution cas coin 2}
\hat{\mu}=m_{-}\quad,\quad
\hat{\lambda}= \left\{
\begin{array}{lcrcl}
   \check{\check{\check{\lambda}}} & \mbox{if}& l_{-}\hspace{-0,6em}&< &\check{\check{\check{\lambda}}}  \leq l_{+} \ , \\
   l_{+}  & \mbox{if}&  l_{+}\hspace{-0,6em} &\leq& \check{\check{\check{\lambda}}} \ ,
\end{array} \ ,
\right.
\end{equation}
and problem is solved.
 \item The case $g_{\mu} < 0$ and $g_{\lambda} < 0$ cannot happen.
\end{itemize}
\end{enumerate}
The minimum value $\widetilde{S}_{\mbox{\tiny min}}$ of $\widetilde{S}$ over $Q$ is then:
\begin{equation}
\label{eq:Shat}
\widetilde{S}_{\mbox{\tiny min}} = A_2-2A\,\hat{\alpha}+\hat{\alpha}^2 + B_2-2B\,\hat{\beta}+\hat{\beta}^2 ,
\end{equation}
where
$ \hat{\alpha}=\hat{\mu}-\hat{\lambda}\ \mbox{ and} \ \hat{\beta}=\hat{\mu}+\hat{\lambda}.$

\end{document}